\documentclass[namedreferences]{solarphysics}
\usepackage{color}
\usepackage[optionalrh,solaromanenum]{spr-sola-addons} 
\usepackage{graphicx}                    
\usepackage{color}                       
\usepackage[normalem]{ulem}
\usepackage{url}                         

\begin{document}

\begin{article}

\begin{opening}

\title{Temporal Evolution of Sunspot Areas and Estimation of Related Plasma Flows}

%
\author{R.~\surname{Gafeira}$^{1,2}$\sep
        C.C.~\surname{Fonte}$^{3,4}$\sep
        M.A.~\surname{Pais}$^{1,5}$
        J.~\surname{Fernandes}$^{1,2,4}$      
       }

%
\runningauthor{R. Gafeira \textit{et al.}}
\runningtitle{Temporal Evolution of Sunspot Areas and Estimation of Related Plasma Flows}

%
  \institute{$^{1}$ Center for Geophysics of the University of Coimbra, Portugal
                     email: \url{gafeira@mat.uc.pt}\\ 
                  $^{2}$ Astronomical Observatory of the University of Coimbra, Portugal\\
             $^{3}$ Institute for Systems and Computers Engineering at Coimbra, Portugal\\
		$^{4}$ Department of Mathematics, University of Coimbra, Portugal\\
		$^{5}$ Department of Physics, University of Coimbra, Portugal
             }

\begin{abstract}
The increased amount of information provided by ongoing missions such as the \textit{Solar Dynamics Observatory} (SDO) represents a great challenge for the understanding of  basic questions such as the internal structure of sunspots and how they evolve with time. Here, we contribute with the exploitation of new data, to provide a better understanding of the separate growth and decay of sunspots, umbra and penumbra. Using fuzzy sets to compute separately the areas of sunspot umbra and penumbra, the growth and decay rates for active regions NOAA 11117, NOAA 11428, NOAA 11429, and NOAA 11430 are computed, from the analysis of intensitygrams obtained by the \textit{Helioseismic and Magnetic Imager} onboard SDO. A simplified numerical model is proposed for the decay phase, whereby an empirical irrotational and uniformly convergent horizontal velocity field interacting with an axially symmetric and height invariant magnetic field, reproduces the large-scale features of the much more complex convection observed inside sunspots.
\end{abstract}

%

\end{opening}

%
 \section{Introduction}\label{Introduction} 

The emergence of magnetic field through the photosphere has multiple manifestations, and sunspots are the most prominent examples of these. Since the discovery of the solar cycle by \inlinecite{1844AN.....21..233S}, sunspots have been extensively studied from very different and complementary perspectives. For instance, we point out recent studies of the sub-photosphere structure using magnetohydrodynamics (MHD) simulations (\textit{e.g.} \opencite{2010ApJ...720..233C}) or the analysis of local helioseismology (\textit{cf}. \inlinecite{2010arXiv1010.4927K} and references therein). See also \inlinecite{2010SoPh..267....1M} for a global discussion on the subject. On the other hand, the sunspot number is a standard parameter used in long time-series analysis of the solar activity in the framework of space weather studies (\textit{e.g.} \opencite{2009SSRv..147..121M}).
During the last four decades, a great amount of data has been collected and several models have been proposed to explain the structure and evolution of sunspots (\opencite{2003A&ARv..11..153S} and references therein). One of the most relevant sunspot properties, to study both its structure and evolution, is the sunspot area: either total, umbra, or penumbra area. The sunspot area has a considerable impact on the solar activity, namely on the variation of the total radiance and the magnetic flux \cite{1997SoPh..175..197P}. Moreover, both total and umbra areas are proxies of the sunspot magnetic field strength \cite{2006Ap&SS.306...23J}. Accordingly, one can find a considerable amount of work dedicated to the study of the characteristics and evolution of sunspot areas. Some of the most important lines of research are:  the analysis of sunspot growth and decay \cite{2006SoPh..237..321C}; the balance between the umbral and penumbral areas and its relation to the total sunspot brightness \cite{2007A&A...465..291M}; the stability of the total umbral area from one cycle to another, as supported by the analysis of long time series covering several solar cycles \cite{1988ApJ...327..451B}. Due to its relevance, large databases of sunspot areas have been built by different groups during the last years (\textit{cf}. \opencite{2005SoPh..228..361Z}, \opencite{2009JGRA..11407104B}) to support the above studies. Simultaneously, a considerable effort has been applied on the ARs automatic detection using different approaches and methods (\textit{e.g.} \opencite{2011A&A...533A..14W}; \opencite{2011SoPh..tmp..369V})

Modern solar instrumentation allows the analysis of sunspot evolution within timescales as short as one minute and even one second. As a result, the evolution of sunspots (areas) can be followed in detail. \inlinecite{2010AN....331..563S} studied the evolution of NOAA AR 11024 during a 4:40 time period, using data in the G-band and Ca {\small II} K from the German Vacuum Tower Telescope. They concluded that during the penumbral formation, the umbral area remains constant and the increase of the total sunspot area is caused exclusively by the penumbral growth. The penumbra region is where the inclination of the magnetic field lines with respect to the direction of local gravity exceeds a critical value of about 30$^\circ$. \inlinecite{2010AN....331..563S} observed that during sunspot growth penumbral filaments begin to form at the umbral boundary, but the umbral area is, basically, invariant with time. On this subject it is important to point out that recent observation of the three-minute oscillation over sunspot umbra confirms a variation of the magnetic field inclination inside the umbra, from 0$^\circ$ at the center to about 30$^\circ$ at the boundary \cite{2012ApJ...746..119R}. Additionally, as pointed out by \inlinecite{2011SoPh..270..463J},``the studies of growth and decay of sunspots or sunspot groups are important for understanding configuration and topology of the magnetic structure on the solar surface, the solar variability and the underlying mechanism of it.''

It is plausible that sunspot evolution is intimately related to the plasma flow in the vicinities of the photosphere. Different techniques to track horizontal proper motions on the photosphere and subphotosphere have recently come forth, such as feature tracking (FT) and local correlation tracking (LCT) methods to estimate surface plasma velocities (\textit{e.g.} \opencite{2009A&A...504..575S}, \opencite{2011A&A...529A.153V}), and helioseismic techniques to assess subphotospheric flows (\textit{e.g.} \opencite{2009ApJ...698.1749H}, \opencite{2010ApJ...708..304Z}). \inlinecite{2012SoPh..tmp..195L} use the two methods to compare horizontal flow fields in the photosphere and subphotosphere. Understanding the observed features of sunspots as accurately as possible is a useful exercise for a more profound understanding of the solar surface/subsurface dynamics. Some recurrent observational results concerning plasma flow and sunspots are i) conspicuous penumbral Evershed outflows of magnitude 1 -- 4 km s$^{-1}$, observed since 1909 and explained by recent numerical simulations \cite{2009ApJ...700L.178K}; ii) outgoing moat flows beyond the sunspots penumbra (\textit{e.g.} \opencite{2012AN....333..125B}, \opencite{2012SoPh..tmp..195L}); iii) inflows in the inner penumbra and umbra, detected with FT and LCT methods (\textit{e.g.} \opencite{2009A&A...504..575S}, \opencite{2011A&A...529A.153V}, \opencite{2012SoPh..tmp..195L}, \opencite{2012AN....333..125B}). The convective scenario in which these features coexist is still unclear. \inlinecite{2010SoPh..267....1M} propose a schematic flow structure of the large-scale circulations within active regions, with a mean inflow at the active regions periphery and a stronger outflow closer to the surface at the core of AR (sunspots). Recent work by \inlinecite{2012SoPh..tmp..195L} using FT and helioseismic methods, compares horizontal flow fields in the photosphere and in a deeper layer 0.5 Mm below the photosphere in two solar active regions. Their results picture a higher resolution flow structure inside the sunspot, with inward flow inside the sunspot umbra and inner penumbra and outward flow starting inside the penumbra and extending to the areas surrounding the sunspot. They found that the inward-flow area in the sunspot is larger at depth.

Since February 2010, the \textit{Solar Dynamics Observatory} (SDO : \opencite{2012SoPh..275....3P}) has been in orbit monitoring the Sun and taking images with time scales and resolutions never achieved before. \inlinecite{2009SoPh..260...21F} presented a new method to determine sunspot areas based on fuzzy sets. The method allows the penumbral area, umbral area, and their corresponding uncertainties to be determined automatically. This methodology is particularly suitable to use with high-resolution images such as those from \textit{Helioseismic and Magnetic Imager} (HMI : \opencite{2012SoPh..275..229S}).
Combining these observational and methodologic developments, this article aims to analyze in detail the areal evolution of the ARs NOAA 11117, NOAA 11428, NOAA 11429, and NOAA 11430. A simplified MHD kinematic model to explain the area evolution of umbra/penumbra during the decay phase is also proposed.

This article is organized as follows: in Section 2 we describe the methodology to derive the umbral and penumbral areas based on a fuzzy sets approach and we present the obtained results for the evolution of the sunspot areas; in Section 3 we test a diffusion--advection numerical model based on the MHD equations to simulate the decay of one sunspot; the last section is devoted to the discussion of results and conclusions.

 \section{ARs Area Computation}\label{Dataresults}

\subsection{Data}\label{Dataset}

The images treated in this study are intensitygrams obtained by the HMI installed onboard the SDO. A set of images was analyzed for the ARs: 583 images for NOAA AR 11117, obtained from 17:55 on 23 October 2010 to 19:55 on 29 October 2010; 733 images for NOAA AR 11428, from 00:15 on 4 March 2012 to 10:15 on 12 March 2012; 870 images for NOAA AR 11429, from 00:15 on 4 March 2012 to 23:45 on 13 March 2012;  620 images for NOAA AR 11429, from 11:00 on 4 March 2012 to 16:15 on 11 March 2012. The images are in JPEG format and have a resolution of 4096 $\times$ 4096 pixels. In this work we have used the images in JPEG  instead of FITS format and have confirmed that with the fuzzy method the difference in the areas obtained is only 5 \%, which is smaller than the error due to other sources. 
Indeed, even a one pixel uncertainty in sunspot radius will introduce an area error larger than 5 \% and so this does not require the data to be reprocessed with FITS files.

With this option we have significantly reduced the amount of data stored and processed, making the process faster.

\subsection{Fuzzy Areas of Sunspots}\label{fuzzy}

There is not a well-defined criterion to separate the pixels belonging to the umbra from those belonging to the penumbra of sunspots, nor between the pixels belonging to the penumbra and the photosphere, especially in high-resolution images. Differences in assigning pixels to the umbra, penumbra or photosphere influence the results obtained for the area of those regions, and therefore there is uncertainty in these values. One possible option to account for this uncertainty is to use approaches based on fuzzy sets, such as the one proposed by \inlinecite{2009SoPh..260...21F} for HMI continuum images and by \inlinecite{2009A&A...505..361B} for solar extreme ultraviolet (EUV) solar images. Here, we follow \inlinecite{2009SoPh..260...21F} to compute the fuzzy umbra, fuzzy penumbra, and fuzzy sunspots area.

A fuzzy set A, defined in a universal set X, is characterized by a membership function $\mu_A$(x) \cite{Klir:1994:FSF:202684}, which expresses the degree of membership of each element of X to the fuzzy set. The degrees of membership usually take values between zero and one, where zero means no membership and one full membership. In this application fuzzy sets corresponding to the umbra, penumbra, and sunspot are generated, as a function of the radiation intensity registered in each pixel. To compute the degrees of membership to assign to the different intensity values the proposed methodology uses a low-pass smoothing filter to decrease the variability of values of the pixels belonging to these three types of regions visible in high-resolution images. A histogram of the filtered image is then analyzed to determine the range of intensity values corresponding to the transition zones between the umbra and the penumbra, and between the penumbra and the photosphere. These values are used to generate membership functions of the intensity values to the umbra, penumbra and photosphere, which are then used to compute the degrees of membership of each pixel to these regions, based on the pixel intensity value. To determine the fuzzy areas of the fuzzy umbra, penumbra, and sunspot, the fuzzy area operator described by \inlinecite{2009SoPh..260...21F} and \inlinecite{DBLP:journals/gis/FonteL04} is used, which enables the identification of pixels where the assignment to one of the regions (umbra, penumbra, or photosphere) is uncertain and evaluates the influence of these doubts in the area computation. This enables the quantification of uncertainty, and allows, among other things, the computation of maximum and minimum area values for the umbra, penumbra, and photosphere, corresponding, respectively, to the area obtained considering the pixels with degrees of membership to each region larger than zero and equal to one.

\subsection{Results of Group Area Analysis}\label{Groupanalysis}

{The methodology was applied to the data described in the section \ref{Dataset}. In the filtering process, a five by five pixel window was used, considering equal weights for all pixels. The filter ran three times for each image. Figures 1, 2, 3, and 4 show the area values (in millionths of the area of a visible solar hemisphere: $\mu$Hem) of the ARs analyzed, during the sunspot growth and decay, as a function of time.}

\begin{figure} 
\centerline{\includegraphics[width=1\textwidth,clip=]{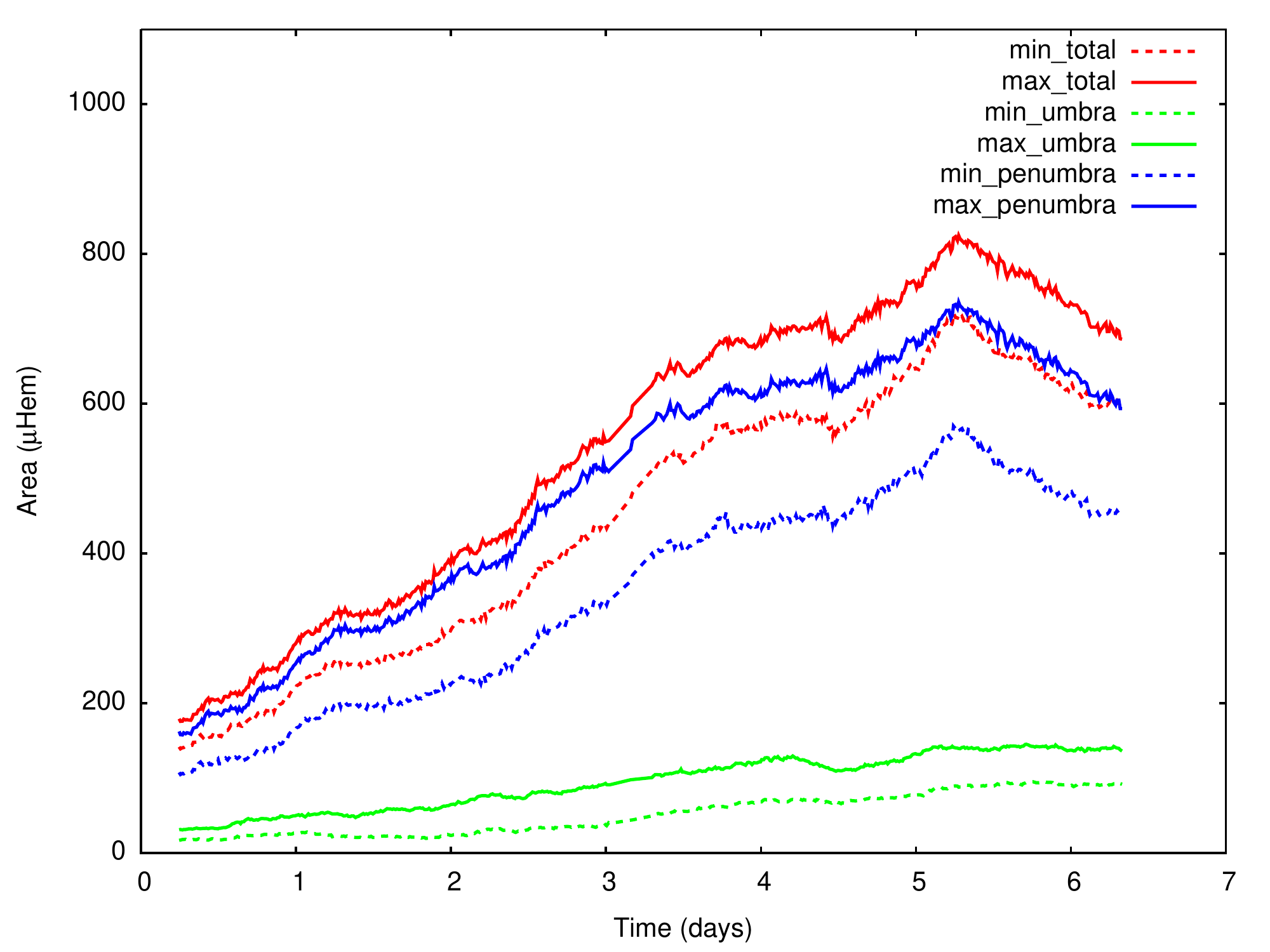}}
\caption{Areal evolution of NOAA AR 11117: minima (dashed lines) and maxima (solid lines) fuzzy area estimates for the total sunspot area (red lines); the umbra (green lines); the penumbra (blue lines).}
\label{fig:11117}
\end{figure}

\begin{figure} 
\centerline{\includegraphics[width=1\textwidth,clip=]{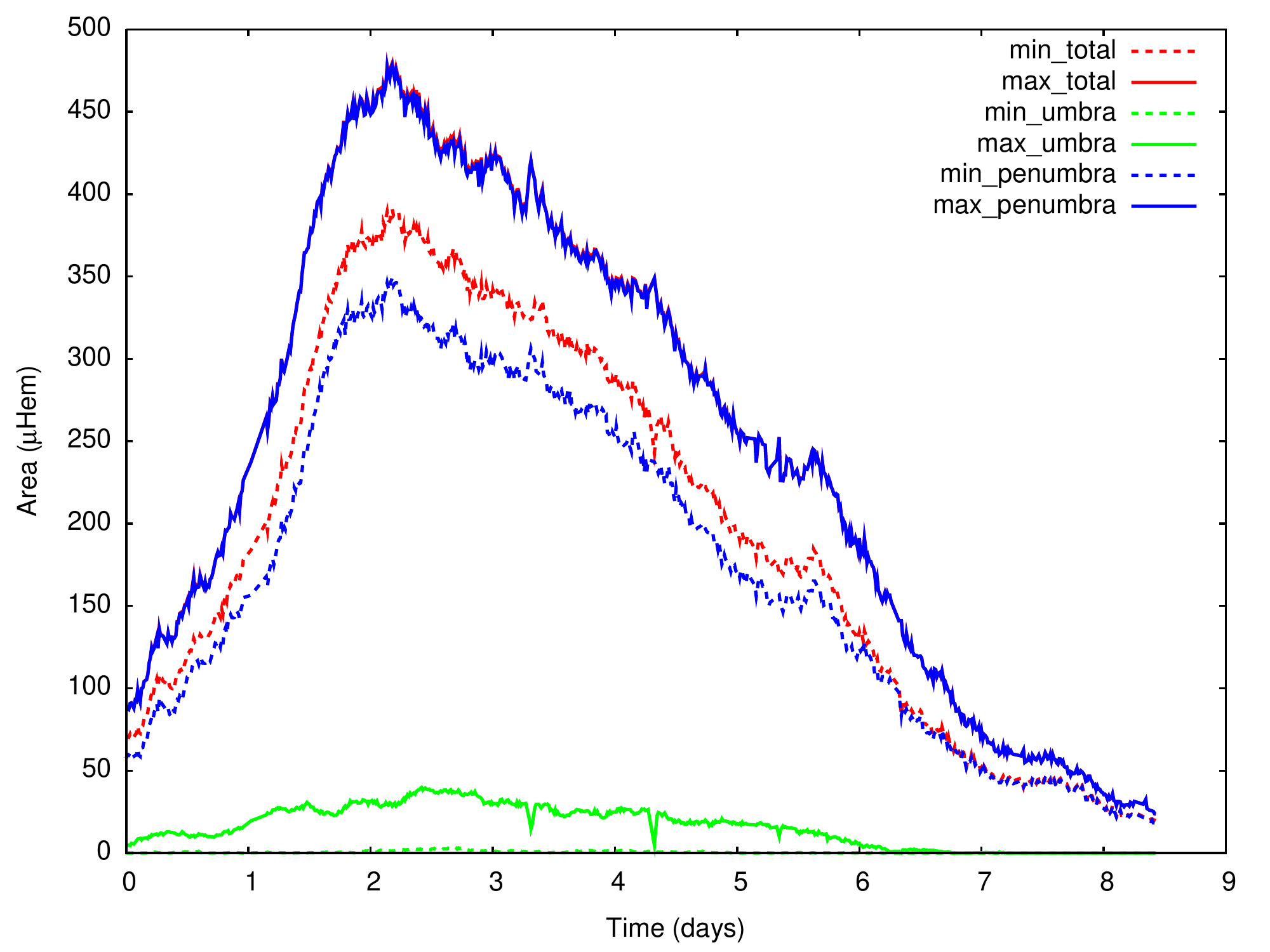}}
\caption{Areal evolution of NOAA AR 11428: minima (dashed lines) and maxima (solid lines) fuzzy area estimates for the total sunspot area (red lines); the umbra (green lines); the penumbra (blue lines).}
\label{fig:11428}
\end{figure}

\begin{figure} 
\centerline{\includegraphics[width=1\textwidth,clip=]{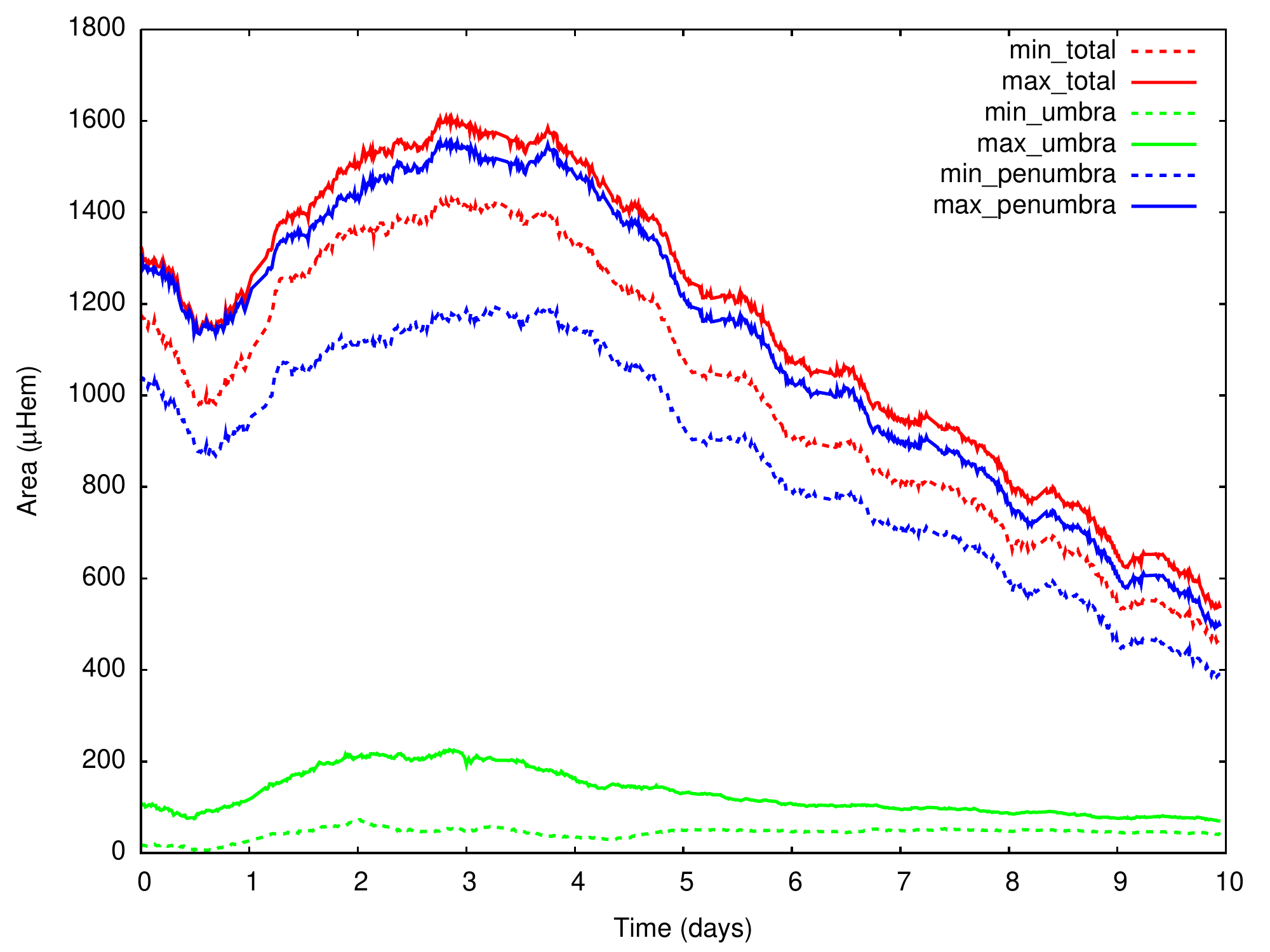}}
\caption{Areal evolution of the NOAA AR 11429: minima (dashed lines) and maxima (solid lines) fuzzy area estimates for the total sunspot area (red lines); the umbra (green lines); the penumbra (blue lines).}
\label{fig:11429}
\end{figure}

\begin{figure} 
\centerline{\includegraphics[width=1\textwidth,clip=]{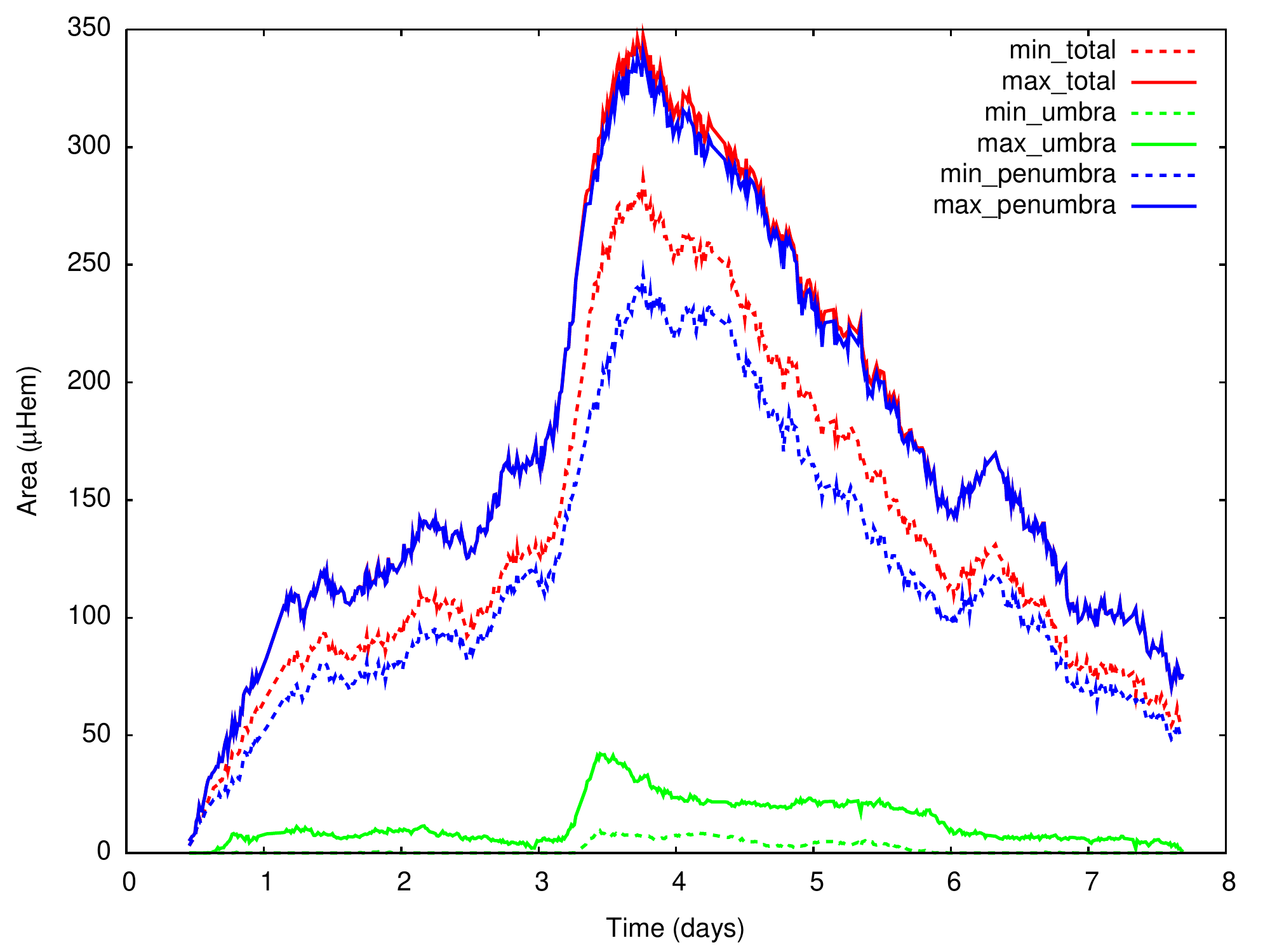}}
\caption{Areal evolution of the NOAA AR 11430: minima (dashed lines) and maxima (solid lines) fuzzy area estimates for the total sunspot area (red lines); the umbra (green lines); the penumbra (blue lines).}
\label{fig:11430}
\end{figure}

Area growth and decay rates (d\textit{A}/d\textit{t}) were computed for each sunspot feature (umbra, penumbra and total sunspot) by fitting a linear regression model to the area evolution curves shown in Figures 1 to 4. Using the maximum and minimum values obtained for the fuzzy area of the umbra, penumbra, and total sunspot (shown in Figure 1), two different estimates are obtained for the increase or decrease rate of these regions. The difference between the two values provides an estimate of the uncertainty associated to the increase/decrease rate. The results are shown in Tables 1 and 2. The values obtained are of the same order of magnitude as those presented by \inlinecite{2008SoPh..250..269H}, who found a decay rate of d\textit{A}/d\textit{t} = -150 $\mu$Hem day$^{-1}$ for AR 9415 one day after the maximal area value was attained. They also confirm the observations of \inlinecite{2010AN....331..563S}, which indicate that the growth and decay of penumbra is the largest contributor to the total area evolution and that umbral growth and decay rates are much smaller.

.

\begin{table}
\caption{Growth phase of AR features areas}
\label{tabgrow}
\begin{tabular}{ccccc}     
\hline
NOAA&Min/max values &Min/max rate values&Min/max rate&Min/max rate \\
 AR &total area&global sunspot &values penumbra &values umbra \\

\, &[$\mu$Hem]&[$\mu$Hem day$^{-1}$]&[$\mu$Hem day$^{-1}$]&[$\mu$Hem day$^{-1}$]\\

\hline
11117&720/820&113/130&90/115&14/22\\
11428&390/480&165/201&148/201&0/12\\
11429&1420/1600&181/196&123/180&16/56\\
11430&295/350&267/300&199/283&16/68\\

\hline
\end{tabular}
\end{table}

\begin{table}
\caption{Decay phase of AR features areas}
\label{tabdecay}
\begin{tabular}{ccccc}     
\hline
NOAA&Min/max values &Min/max rate values&Min/max rate&Min/max rate \\
AR &total area&global sunspot &values penumbra &values umbra \\

&[$\mu$Hem]&[$\mu$Hem day$^{-1}$]&[$\mu$Hem day$^{-1}$]&[$\mu$Hem day$^{-1}$]\\

\hline
11117&720/820&-116/-126&-112/-127&1/-1\\
11428&390/480&-77/-85&-66/-85&-0/-8\\
11429&1420/1600&-121/-142&-108/-141&-1/-10\\
11430&295/350&-71/-84&-65/-81&-3/-4\\


\hline
\end{tabular}
\end{table}

%


 \section{Simulation of Sunspot Area Decay}\label{Simulation} 
We will focus our analysis on the sunspot decay phase which, as pointed out by different authors (\textit{e.g.} \opencite{2008ApJ..679..676}) is less understood than penumbral formation and growth. It is assumed that the main features of the umbral and penumbral decay phases can be captured with the approximation of cylindrical symmetry relative to the sunspot axis and a negligible height dependence. Denoting cylindrical coordinates by $(r, \theta, z)$, where \textit{r} is the distance to the axis, $\theta$ is the azimuthal angle and $z$ is the height above the photosphere, then all partial derivatives with respect to $z$ and $\theta$ are zero. The first step was to check if the sunspot behavior during the decay phase could be an effect of diffusion, either ohmic or turbulent. The system of equations describing diffusion of a poloidal magnetic field with the assumed symmetry is given by


\begin{equation}
\label{difz}
\frac{\partial B_z}{\partial t} = \eta \nabla^2 B_z
\end{equation}

\begin{equation}
\label{difr}
\frac{\partial B_r}{\partial t} =\eta \left( \nabla^2 B_r - \frac{B_r}{r^2} \right)
\end{equation}
where $B_z(r,t)$ and $B_r(r,t)$ are the axial and radial magnetic field components, $t$ is time, and $\eta$ is the magnetic diffusivity. The $\nabla^2$ operator is given by $$\nabla^2=\frac{1}{r}\frac{\partial}{\partial r} \left( r \frac{\partial}{\partial r} \right).$$

\begin{figure}    
\centerline{\includegraphics[width=0.8\textwidth,clip=]{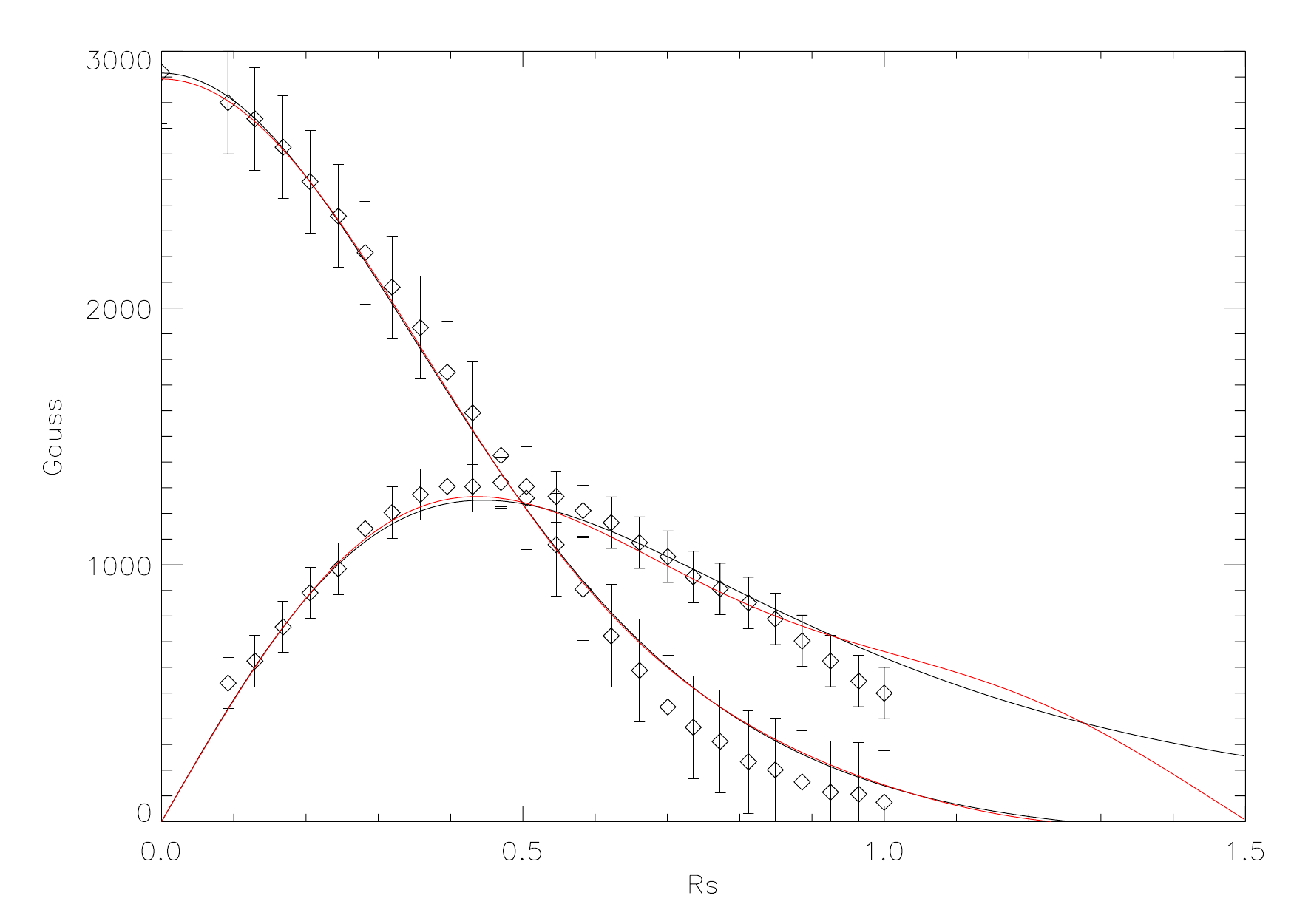}}
\caption{For the sunspot magnetic field components $B_z$ and $B_r$ of AR 10933: radial model (\protect\inlinecite{2011LRSP....8....4B}, white diamonds) with error bars, best buried-dipole model (\protect\inlinecite{2003A&ARv..11..153S}, black lines) and the model proposed in this article of free decay cylindrical modes (red lines).}
\label{campplot}
\end{figure}

The differential Equations (\ref{difz}) and (\ref{difr}) can be analytically solved using separation of variables. As boundary conditions, we require that both $B_z$ and $B_r$ must be finite at the sunspot axis (r = 0). The outer boundary conditions, as well as the initial conditions [$B_z(r/r_0,0)$ and $B_r(r/r_0,0)$] are obtained from \inlinecite{2011LRSP....8....4B}, who computed the radial variation of azimuthally averaged vertical and horizontal components of the magnetic field from the inversion of spectropolarimetric observations employing a Milne--Eddington atmospheric model. We found for the two initial radial functions [$B_z(r/r_0,0)$ and $B_r(r/r_0,0)$] a combination of Bessel functions of the first kind respectively of orders 0 and 1. These initial functions, which very closely represent the \inlinecite{2011LRSP....8....4B} model for AR 10933 (see Figure 11 in \inlinecite{2011LRSP....8....4B}) as shown in Figure \ref{campplot}, are expressed by Equations (\ref{bessel_bz}) and (\ref{bessel_br})

\begin{eqnarray}
B_z(r/r_0)& = &1166.51\ J_0(k_{1,0} r/r_0) + 1157.55\ J_0(k_{2,0} r/r_0) \nonumber \\
&& + 382.855\ J_0(k_{3,0} r/r_0) + 162.414\ J_0(k_{4,0} r/r_0) \label{bessel_bz}
\end{eqnarray}

\begin{eqnarray}
B_r(r/r_0) & = & 1688.57\ J_1(k_{1,1} r/r_0) + 530.185\ J_1(k_{2,1} r/r_0) \nonumber \\
&&+ 428.751\ J_1(k_{3,1} r/r_0) \label{bessel_br}
\end{eqnarray}
where $r_0 = 1.5R_S$ with $R_S$ the initial sunspot radius, $k_{m,0}$ is the $m$-th root of $J_0(r/r_0)$, and $k_{m,1}$ is the $m$-th root of $J_1(r/r_0)$. According to \inlinecite{2011LRSP....8....4B}, the umbra/penumbra and the penumbra/moat separation conditions are $B_r/B_z = $tan$\ 35^\circ$ and
$B_r/B_z = $tan$\ 77^\circ$, respectively. Based on Figure 11 in \inlinecite{2011LRSP....8....4B} we use variance values of $\sigma^2_{B_z}	= 100^2 G^2$ and $\sigma^2_{B_r}	= 200^2 G^2$ for the fit of our Equations (\ref{bessel_bz}) and (\ref{bessel_br}) to the \inlinecite{2011LRSP....8....4B} radial model. We obtain $\chi^2$ values $\chi^2_{B_z}	= 0.006$ and $\chi^2_{B_r}	= 0.07$, respectively, for the  two fits. Our initial solution is also very close to the buried-dipole solution \cite{2003A&ARv..11..153S} for a dipole magnetic moment of $1.25 \times 10^{23}$ Am$^2$ and a dipole depth of $20\ 470$ km.

The diffusion process can accordingly be described in terms of the exponential decay of just a few modes, each one evolving with a characteristic time-scale:

\begin{equation}
\label{beselz}
B_z(r/r_0,t)=\sum^{4}_{m=1}A_mJ_0(k_{m,0}r/r_0)\exp(-\eta k^2_{m,0}t/r^2_0)
\end{equation}

\begin{equation}
\label{beselr}
B_r(r/r_0,t)=\sum^{3}_{m=1}A_mJ_1(k_{m,1}r/r_0)\exp(-\eta k^2_{m,1}t/r^2_0)
\end{equation}

The important point to realize here is that the temporal evolution of the sunspot geometry during a purely diffusive process is totally determined by the relative contribution of different modes for $B_z(r/r_0)$ and $B_r(r/r_0)$. In particular, the fact that the most important mode in $B_z(r/r_0)$ decays with a time scale of $r^2_0/(k^2_{1,0}\eta)$ where $k^2_{1,0}$= 5.78, longer than the dominant time scale $[r^2_0/(k^2_{1,1}\eta)]$ of $B_r$ decay where $k^2_{1,1}$=14.68, makes the inclination of magnetic field lines at a certain point inside the sunspot decrease with time, and as a result the umbral area increases with time. This result, which is independent of the value adopted for the magnetic diffusivity as long as it is the same for both Equations (1) and (2), is incompatible with observations.

Inward advection of the magnetic field lines by the plasma flow can counteract the effect of diffusion, for adequately chosen velocity values. In order to simulate this effect, a simplified kinematic model was tested whereby the advection and stretching of the magnetic field by the photospheric plasma is the main mechanism responsible for the sunspot global area decay. In the approximation of a radial sunspot, the plasma velocity field is of the form $\textbf{u} = u_r(r)r$ and the induction equations to solve are (\ref{indz}) and (\ref{indr}):



\begin{eqnarray}
\frac{\partial B_z}{\partial t} & = &-u_r\frac{\partial B_z}{\partial r}+ \eta \nabla^2 B_z \label{indz}
\\
\frac{\partial B_r}{\partial t} & = &B_r \frac{\partial u_r}{\partial r}- u_r\frac{\partial B_r} {\partial r}+ \eta \left( \nabla^2 B_r - \frac{B_r}{r^2} \right)\label{indr}
\end{eqnarray}

Furthermore, as the observed decrease of penumbra area is steeper than the observed umbral area decrease, we can anticipate a higher inward flow affecting the penumbra than that affecting the umbra. The velocity field made to interact with the magnetic field is radial of the form $u_r = u_0r/R_S$, where $r$ is the radial coordinate and $R_S$ is the initial sunspot radius. In order to simulate the diffusion effect, an ohmic diffusivity of $\eta = 0.2$ km$^2$ s$^{-1}$ was used, as estimated by \inlinecite{2008ApJ...683.1153C}.

\begin{figure}    
  \centerline{\includegraphics[width=1\textwidth,clip=]{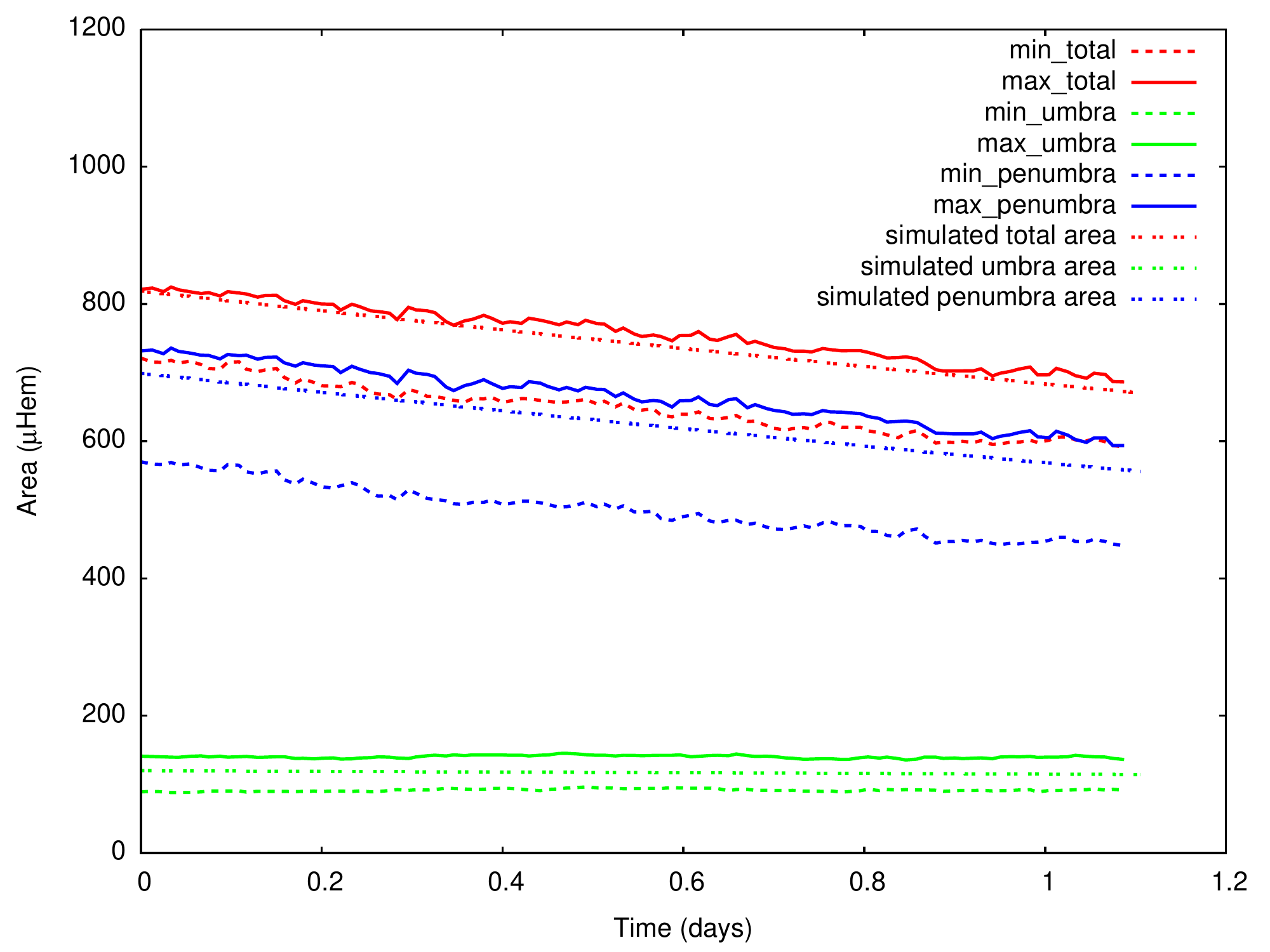}}
  \caption{Decay phase of areal evolution for NOAA AR 11117 and corresponding simulated curves. Simulated values (intermittent dashed lines) minima fuzzy estimates (dashed lines) and maxima fuzzy estimates (solid lines) for the total sunspot area (red lines); the umbra (green lines); the penumbra (blue lines).}

   \label{figressimdec}
   \end{figure}

For each of the four active regions NOAA 11117, 11428, 11429, and 11430, the $u_0$-parameter was adjusted in order that the linear fit to the simulated sunspot decay could very closely reproduce the linear fit to the observed sunspot decay (see Table \ref{tabdecay}). The result for AR 11117 is shown in Figure \ref{figressimdec}. Table \ref{tab3} shows the numerical values found for $u_0$, together with initial sunspot areas, [$A_S$], and initial sunspot radius [$R_S$], of different ARs. The two estimates shown for each one of the parameters $u_0$, $A_S$, and [$R_S$] correspond to fits made to minima and maxima fuzzy area curves shown in
Figures 1, 2, 3, and 4. We consider as estimates for the error of $u_0$ $[\sigma_{u_0}]$ the differences between the estimated $u_0$-values using minima and maxima fuzzy areas.

\begin{table}

\label{tabvelo}
\begin{tabular}{cccccc}     
\hline
AR NOAA&$\Delta t$&$A_S$&$R_S$&$u_0$&$\sigma_{u_{0}}$\\
&[days]&[$\mu$Hem]&[Mm]&[m s$^{-1}$]&[m s$^{-1}$]\\
\hline
11117&1.5&720/820&26.41/27.84&-34.6/-35.3&0.7\\
11428&4.0&390/480&19.44/21.57&-59.0/-48.0&11.0\\
11429&6.0&1420/1600&37.09/39.37&-32.0/-35.0&3.0\\
11430&2.5&295/350&16.01/18.42&-44.0/-48.0&4.0\\
\hline
\caption{Fitted values of $u_0$ (minimum/maximum) for the four case studies considered. Also shown are the time length used for estimating the linear decay rate [$\Delta t$], the initial (minimum/maximum) sunspot area [$A_S$], the initial (minimum/maximum) sunspot radius [$R_S$] and the error estimation associated to the choice of $u_0$ [$\sigma _{u_0}$].}
\label{tab3}
\end{tabular}
\end{table}

 \section{Discussion and Conclusions}\label{Discussionconclusions}

Recently \inlinecite{2010AN....331..563S} studied the evolution of NOAA AR 11024 concluding that during the penumbral formation the umbral area remains constant and that the increase of the total sunspot area is caused exclusively by the penumbral growth.

The results obtained in this study for areal evolution using fuzzy areas estimation seem to not contradict the conclusions of \inlinecite{2010AN....331..563S}.
In fact, taking into account the short temporal duration of the observations in their study (only 4.7 hours), it is possible that the umbral growth could not be perceived if, as we see here, its growth rate was much weaker than that of the penumbra.


A numerical kinematic model was tested, whereby the advection and stretching of the magnetic field by the photospheric plasma is the main mechanism responsible for the sunspot area decay. A \textit{z}-independent initial solution for the sunspot's magnetic field was used, in agreement with the underlying assumptions. The obtained results seem to meet previous claims that the advection of field due to an inflow inside an active region with speeds between 10 and 100 m s$^{-1}$ should be required to balance the outward transport of magnetic field by turbulent diffusion \cite{1538-4357-684-2-L123}. According to Hurlburt and DeRosa's numerical simulations of compressible magnetoconvection with flux-dependent surface cooling, the inflow inside the active regions is driven by buoyancy as a result of localized surface cooling inside active regions. Our estimates for the maximum speed $u_0$ occurring at the penumbra periphery, with values lying between 32 -- 59 m s$^{-1}$ for the case studies considered, does in fact match the order of magnitude of \inlinecite{1538-4357-684-2-L123} large-scale inflows.  Our results are not incompatible with the well-known Evershed outflows, since as shown by \inlinecite{2009ApJ...700L.178K} low magnitude inflows of umbral and penumbral features can coexist with Evershed outflows as part of overturning convection motions. The authors expect that, thanks to the high-quality instrumentation onboard of SDO or \textit{Hinode} and to existing tools from helioseismology and photospheric feature tracking, it will be possible in the future to test the numerical predictions presented in this article for a large number of active regions. 

%

%
\begin{acks}
The authors wish to thank NASA/SDO and the HMI science team for providing the HMI images and the reviewer for constructive comments that helped to improve the manuscript.MAP is supported by FCT (PTDC/CTE-GIX/119967/2010) through the project COMPETE (FCOMP-01-0124-FEDER-019978).CFC's work was partially supported by Funda\c{c}\~ao para a Ci\^encia e a Tecnologia (FCT) under project grant PEst-C/EEI/UI0308/\-2011.
\end{acks}

%
%
 \bibliographystyle{spr-mp-sola}
%

%
%
%

\end{article} 
\end{document}